\documentclass[10pt,letterpaper,twocolumn]{article} %% two column, final layout
\usepackage{ol2}
\usepackage[draft]{hyperref}
\usepackage{amsmath}
\usepackage[applemac]{inputenc}

\begin{document}

\twocolumn[ %% activate for two-column option

\title{From higher-order Kerr nonlinearities to quantitative modeling of 3rd and 5th harmonic generation in argon}

%% For REVTeX it is possible to automate superscript and e-mail callouts with the superscriptaddress option; see REVTeX4 documentation.

\author{P.  B\'ejot,$^{1}$ E. Hertz,$^1$ B. Lavorel$^1$, J. Kasparian$^2$, J.-P. Wolf$^2$, and O. Faucher$^{1,*}$}

\address{
$^1$Laboratoire Interdisciplinaire CARNOT de Bourgogne (ICB), UMR 5209 CNRS-Universit\'e de Bourgogne\\ BP 47870, 21078 Dijon Cedex, France
\\
$^2$
Universit\'e de Gen\`eve, GAP-Biophotonics, 20 rue
de l'Ecole de M\'edecine, 1211 Geneva 4, Switzerland
 \\

%$^*$Corresponding author:olivier.faucher@u-bourgogne.fr
}
\begin{abstract}
The recent measurement of negative higher-order Kerr effect (HOKE) terms
in gases has given rise to a controversial debate, fed by its  impact on short laser pulse propagation. By comparing the experimentally measured yield of the third and fifth harmonics,
with both an analytical and a full comprehensive numerical propagation model, we confirm the absolute and relative values of the reported HOKE indices.
\end{abstract}


 ] %% activate for two-column option

\noindent In a recent experiment, we have  shown that the electronic optical Kerr effect in Ar, N$_2$, O$_2$, and air exhibits a highly nonlinear behavior versus the applied intensity \cite{Loriot}, resulting in a saturation of the nonlinear refractive index observed at moderate intensity, followed by a sign inversion at higher laser intensity.  This observation has a substantial impact on the propagation of  ultrashort and ultra-intense laser pulses, especially in the context of laser filamentation \cite{ChinHLLTABKKS2005,BergeSNKW2007,CouaiM2007,KaspaW2008} where the higher-order Kerr effect (HOKE), rather than the defocusing contribution of the free electrons, can play a key role in the self-guiding process \cite{Bejot_2010}, especially at long wavelengths \cite{Ettoumi} and for short pulses \cite{LorioBEPKHHLFW2010}. However, this issue is still controversial \cite{Kolesik,PolynKWM2010,KolesMDM2010}. Therefore, an independent confirmation of our measurement of the HOKE is still needed.  Recently, Kolesik \textit{et al.} \cite{Kolesik} have proposed such test, based on the comparison of the yields of the third harmonic (TH) and  the fifth harmonic (FH) radiations generated by the nonlinear frequency up-conversion of a short and intense laser pulse in air. Based on numerical simulations, they suggested that, considering the HOKE indices, ``the relative strength of the FH to the TH should reach values of the order of 10$^{-1}$'' while, if omitting them, ``this ratio should be about 4-5 orders smaller'' \cite{Kolesik}.

So far, no  measurement of  the yield of FH versus the TH have been achieved in air. However, Kosma \textit{et al.}\cite{Kosma} measured the yields of TH and FH produced by  a short and intense laser pulse in argon. The present paper aims at confronting the results of this experiment to predictions based on the HOKE in argon\cite{Loriot}.

In the first part, we confirm the ratio of the recently measured non-linear indices \cite{Loriot} based on the analytical description  of the harmonic generation. In the second part, a comprehensive model including linear and nonlinear propagation effects such as dispersion, self-phase modulation, ionization, and Kerr effect, is presented.

For a focused laser beam propagating linearly, the harmonic power of the $q$th harmonic in the perturbative regime is given by
\begin{equation}
P_q=A_qN^2\left|J_q(b\Delta k)\right|^2,
\label{harmopower}
\end{equation}
where  $N$ is the atomic density of the medium and
\begin{equation}
A_q=\frac{q\omega_1^2}{4n^\ell_q (n^\ell_1)^q(\epsilon_0\pi)^{q-1}c^q w_0^{2q-2}}\left(\chi^{(q)}\right)^2 P_1^q,
\label{harmamp}
\end{equation}
with $P_1$,  $\omega_1$, and $w_0$  the power,  the angular frequency, and  the beam waist  of the  incident beam, respectively \,\cite{Boyd,Reintjes}. $\chi^{(q)}$ is the $q$th-order microscopic nonlinear susceptibility ($q=3, 5$) given in SI units, $n^\ell_j$ are the linear refractive indices at the fundamental ($j=1$) and harmonic frequencies ($j=3,5$),
$\epsilon_0$ is the permittivity of vacuum, and $c$ is the speed of light. $J_q$  is a dimensionless function that accounts for the phase matching
\begin{equation}
J_q=\int_{-2f/b}^{2(L-f)/b}\mathrm{d}\xi\frac{\exp\left(-i  b\Delta k\xi /2\right)}{\left(1+i\xi\right)^{q-1}},
\label{phasematch}
\end{equation}
with $\Delta k=k_q-qk_1=\frac{2\pi q}{\lambda_1}\left(n^\ell_q-n^\ell_1\right)$ the phase mismatch, with  $n^\ell_q-n^\ell_1$ proportional to the pressure, and $k_j$ ($j=1, q$) the wave vectors, $b$ the confocal parameter, $L$ the length of the static cell, and $f$ the position of the focus with respect to the entrance of the static cell\,\cite{Bjorklund}. According to Eqs.\,(\ref{harmopower}) and (\ref{harmamp}), the ratio of the FH to the TH power is
\begin{equation}
\frac{P_5}{P_3}\approx\frac{5}{3\epsilon_0^2\pi^2c^2w_0^4}\left(\frac{\chi^{(5)}}{\chi^{(3)}}\right)^2 \left(\frac{N_5\left|J_5\right|}{N_3\left|J_3\right|}\right)^2 P_1^2,
\label{powerratio}
\end{equation}
where $n^\ell_j$  have been approximated  to unity in Eq.\,\ref{harmamp}. $N_3$ and  $N_5$ refer to the different atomic densities  at the pressures maximizing the  harmonic conversion for the 3rd and 5th orders, respectively. This equation provides a direct relationship between the power ratio of the harmonics and the ratio of the corresponding non-linear susceptibilities. The latter are related to the nonlinear refractive indices through the relation\,\cite{Ettoumi}
\begin{equation}
\label{indexversussusc}
n_{2j}=\frac{\left(2j+1\right)!}{2^{j+1}j!\left(j+1\right)!}\frac{1}{\left((n^\ell_1)^2\epsilon_0c\right)^j}\chi_{\mathrm{Kerr}}^{(2j+1)}.
\end{equation}
so that
\begin{equation}
\frac{P_5}{P_3}\approx\frac{3}{5 \pi^2w_0^4}\left(\frac{n_4}{n_2}\right)^2 \left(\frac{N_5\left|J_5\right|}{N_3\left|J_3\right|}\right)^2 P_1^2,
\label{powerratio_n_i}
\end{equation}

\begin{figure}[tb]
\centerline{
\includegraphics[width=8.3cm]{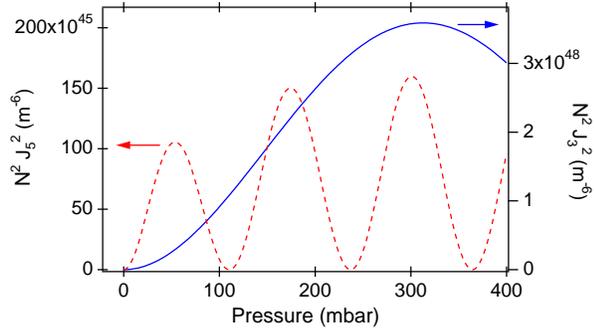}}
 \caption{\label{fig1} (Color online) Analytical calculation of the pressure dependence  of the 3rd (solid blue line) and 5th (dashed red line) harmonics in argon.}
\end{figure}

In the experiment by Kosma \textit{el al.},  $b$=7.8\,cm, $w_0=100\,\mu$m, $L$=1.8\,cm, $f$=$L/2$, and the wavelength $\lambda_1$=810\,nm \cite{Kosma}. The fundamental power, calculated from the input energy $E_1$=710\,$\mu$J and the pulse duration $\tau_1$=12\,fs, is $P_1=59$\,GW. They observed that the pressure maximizing the TH power ranged between 160\,mbar \cite{Kosma} and 250 mbar \cite{Kosma2},  for similar experimental conditions. One single maximum, around  50\,mbar, is observed for the FH. The maximum  energies of the TH and FH measured at the respective optimal pressures reported in  \cite{Kosma} are 140 and 4\,nJ, respectively, while the pulse duration was estimated to be 11\,fs for both harmonics \cite{Kosma,Kosma2}. This leads to a power ratio $P_5/P_3=0.028$.
According to Eq. \,(\ref{powerratio_n_i}), where $J_q$ of Eq.\,(\ref{phasematch}) has been  calculated using  $n^\ell_1=1.00028$, $n^\ell_3=1.00030$, and $n^\ell_5=1.00035$ for the values of the  refractive index of argon at 1 bar at 810, 270, and 162\,nm, respectively\cite{Bideau_1981}, the corresponding ratio of the HOKE indices is $\left|n_4/n_2\right| = 6.8\times10^{-19}$\,m$^2$/W. This value confirms, within a factor of 2  compatible with the experimental error, the ratio of the experimental HOKE indices $n_2=10^{-23}\,\mathrm{m}^2/\mathrm{W}^{1}$ and $n_4=-3.6\times 10^{-42}\,\mathrm{m}^4/\mathrm{W}^{2}$\,\cite{Loriot}, resulting in  $\left|n_4/n_2\right| = 3.6\times10^{-19}$\,m$^2$/W. The agreement is remarkable, especially considering the simplicity of the analytical model used.

Further comparison with the experiment was performed by computing the value of $N^2\left|J_q\right|^2$ as a function of the argon pressure relying on Eq.\,(\ref{phasematch}) (Fig.\,\ref{fig1}).
This function should reflect the pressure dependence of the harmonic powers. The analytical  model predicts a maximum at about 300\,mbar for TH,  in line with the experimental results. It yields three maxima between 0 and  400\,mbar for FH, the first of them close to the observed optimum pressure for the FH. This oscillatory structure, which is due to the periodic phase matching, was not observed in the experiments \cite{private} probably due to non-linear propagation effects which are not considered in the analytical model.

To overcome these limitations and take into account the perturbations of the fundamental pulse during its propagation through the gas sample, as well as the effect of the HOKE indices on the phase matching,
we have solved the unidirectional pulse propagation equation for the  experimental  conditions of Kosma \textit{et al.}
More precisely, assuming a cylindrical symmetry around the propagation axis $z$, the angularly resolved spectrum $\widetilde{E}(k_\bot,\omega)$ of the \textit{real} electric field $E(r,t)$ follows the equation \cite{Moloney}
\begin{eqnarray}
\partial_z\widetilde{E}=ik_z\widetilde{E}+\frac{1}{2k_z}\left(\frac{i\omega^2}{c^2}\widetilde{\mathcal{P}}_{\mathrm{NL}}-\frac{\omega}{\epsilon_0c^2}\widetilde{\mathcal{J}}\right),
\label{Eq1}
\end{eqnarray}
where $k_z=\sqrt{k^2(\omega)-k^2_\bot}$, $\widetilde{\mathcal{P}}_{\mathrm{NL}}$ (resp., $\widetilde{\mathcal{J}}$) is the angularly resolved nonlinear polarization (resp., free charge induced current) spectrum, and $k(\omega)=\frac{n(\omega)\omega}{c}$.

The nonlinear polarization $P_{\mathrm{NL}}$ is evaluated in the time domain as $P_{\mathrm{NL}}=\chi^{(3)}E^3+\chi^{(5)}E^5+\chi^{(7)}E^7+\chi^{(9)}E^9+\chi^{(11)}E^{11}$. Since the nonlinear polarization is defined  from the real electric field, Eq. \ref{Eq1}  captures  without any modifications all frequency-mixing processes induced by the total field. For numerical stability concerns, we considered only the part responsible for the refractive index change around $\omega_0$, neglecting harmonics generation induced by
the terms proportional to $E^{7}$, $E^{9}$, and $E^{11}$.
The current induced by the free charges is calculated in the frequency domain as $\widetilde{\mathcal{J}}=\frac{e^2}{m_e}\frac{\nu_e+i\omega}{\nu_e^2+\omega^2}\widetilde{\rho\varepsilon}$, where $e$ (resp., $m_e$) is the electron  charge (resp., mass), $\nu_e$ is the effective collisional frequency, and $\rho$ is the electron density which is evaluated as
\begin{equation}
\partial_t\rho=W(I)\left(\rho_{\mathrm{at}}-\rho\right)+\frac{\sigma}{U_i}I-\beta\rho^2,
\end{equation}
where $W(I)$ is the ionization probability evaluated with the Keldysh-PPT (Perelomov, Popov, Terent'ev) model\,\cite{BergeSNKW2007},
$\rho_{\mathrm{at}}$ is the atomic number density, $\sigma$ is the inverse Bremsstrahlung cross-section, $\beta$ is the recombination constant (negligible on the time scale investigated in the present work), and $I$ is proportional to the time-averaged $\left<E^2\right>$.

\begin{figure}[tb]
\centerline{
 \includegraphics[width=8.3cm]{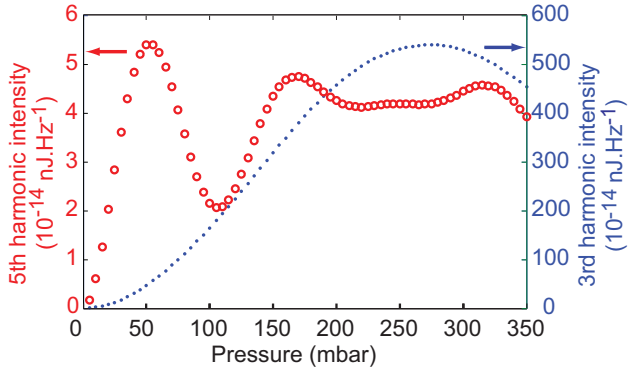}}
  \caption{ \label{fig2} (Color online) Numerical calculation of the pressure dependence of the 3rd (dotted blue line) and 5th (open red circles) harmonics in argon integrated over the full radial distribution. To be compared  with the Fig.\,3 of\,\cite{Kosma}. The spectrum calculated  at 50 mbar is shown in the inset.}
\end{figure}

Figure \ref{fig2} displays the harmonics intensity as a function of argon pressure for an input pulse and a detection geometry matching the experimental parameters: 12 fs pulse duration (FWHM), $700\,\mu$J input energy, and a beam radius of 4 mm before focusing. In order to mimic the experiment, the pulse first propagates in vacuum up to the position of the cell  (99.1 cm after the $f$=1\,m lens). After this focusing step, the pulse propagates over 1.8\,cm in the argon cell. The optimal pressure for the FH is 50 mbar, in full agreement with the experiment \cite{Kosma}. The reduction of the second and third maxima of the FH, as compared to Fig.\,\ref{fig1}, results from the phase mismatch introduced by the HOKE  at large  pressure.  The TH yield is maximal at 260 mbar, similar to the value reported in \cite{Kosma2}. In full agreement with the experiment by Kosma \textit{al.} \cite{Kosma}, the ratio at 50\,mbar is about 0.1 and becomes even larger at reduced  pressures. Furthermore, the total FH and TH energies at their respective optimum pressures are 6 and 218\,nJ, in good agreement with the experimental values  of 4 and  140 \,nJ, where losses due to the setup lead to a slight underestimation of the output energies \cite{Kosma}.

If the HOKE are not considered in the model,
the ratio  of the FH to the TH at a pressure of 50\,mbar drops to 0.017,
and the FH and TH energies are respectively 1.7  and 584\,nJ: These values are inconsistent with the experimental results of Kosma \textit{et al.}
Furthermore, contrary to the experimental observations\,\cite{private}, the FH would exhibit strong maxima at 160 and 250\,mbar. These discrepancies show that the HOKE are necessary to reproduce the experimental results \cite{Kosma,Kosma2}, further validating their measured values \cite{Loriot}.
Note that the ratio of 0.017 strongly depends on the propagation distance, so that it cannot be directly compared to that of 10,000 predicted by Kolesik \textit{et al.} for the “classical” model over an unspecified propagation distance. For a propagation length of 220\,$\mu$m, 80 times shorter than in our work but consistent with neglecting the phase matching, our calculation indeed predicts a ratio of 10,000.

In conclusion, as recently suggested in \cite{Kolesik}, we have compared the recent experimental measurements of the TH and FH yields in argon \cite{Kosma} with both analytical and numerical simulations. These results agree quantitatively with the measured  high-order Kerr indices \cite{Loriot}.  This conclusion  is supported by the following findings. First, the harmonic yield reported  in argon by Kosma \textit{et al.} at the pressure that optimized the 5th harmonic leads to a ratio of about 0.1 between the fifth and the third  harmonics.  This ratio implies a ratio of the Kerr indices consistent with our measurement of the HOKE indices  within their uncertainty range\,\cite{Loriot}. Second, the analytical model based on our HOKE indices reproduces  the pressure maximizing the TH, as well as  the first pressure maximum of  the 5th harmonic yield. Third, a full numerical propagation model accounting for  the dispersion and nonlinear effects such as  ionization and higher-order Kerr effects quantitatively reproduces  the ratio of the harmonic yields observed in the experiment, as well as the pressure dependence of both the 3rd and 5th harmonics. It even reproduces the absolute harmonics intensity within a fairly good accuracy.

\section*{Acknowledgments}
The authors are grateful to Werner Fu{\ss} and Kyriaki Kosma
for fruitful discussions and critical reading of the manuscript.
This work was supported by the Conseil R\'egional de Bourgogne, the ANR \textit{COMOC}, the \textit{FASTQUAST} ITN Program of the 7th FP and the Swiss NSF (contract 200021-125315).

% \bibliography{ref_THG_versus_FHG}
\end{document}